\documentclass[aps,11pt,preprintnumbers,nofootinbib,floatfix]{revtex4}
\usepackage{graphicx}
\usepackage{dcolumn}
\usepackage{bm}
\usepackage{amsmath}
\usepackage{mathtools}
\usepackage{slashed}
\usepackage{amssymb}

\begin{document}

\title{Thermodynamic topology of black holes and an invariant of spacetime}
\author{Cao H. Nam$^{1,2}$}
\email{caohoangnam@duytan.edu.vn}  
\affiliation{$^1$Institute of Theoretical and Applied Research, Duy Tan University, Hanoi 100000, Vietnam\\
$^2$School of Engineering and Technology, Duy Tan University, Da Nang 550000, Vietnam}
\date{\today}

\begin{abstract} 
We present a formalism for exploring the thermodynamic topology of black holes, without introducing the auxiliary variable $\Theta\in[0,\pi]$ employed in previous studies. The formalism is based on an off-shell grand free energy appropriate to thermodynamic systems that can exchange both energy and matter with their environment, thereby allowing thermal and chemical equilibrium. We define a vector field as the gradient of the off-shell grand free energy, whose zeros correspond to equilibrium configurations and hence to black hole solutions, and then we construct a conserved topological tensor from its normalized components. The associated topological flux is given by the sum of local topological indices of all zeros contained within a hypersurface at fixed ensemble parameters. We further define the asymptotic topological flux as the limiting value of the conserved topological flux through a family of finite hypersurfaces as they approach the asymptotic boundary of the ensemble parameter space. We find that the asymptotic topological flux is determined by the asymptotic/background spacetime geometry. This motivates the conjecture that the asymptotic topological flux is an invariant associated with the asymptotic/background spacetime geometry. Within the spacetime classification based on the asymptotic topological flux, the flat-space limit of AdS is found to be nontrivial, providing a potentially complementary perspective on the AdS distance conjecture.
\end{abstract}

\maketitle

\section{Introduction}

Black hole thermodynamics \cite{Bekenstein1973,Hawking1975} has been an active research area in modern physics because it reveals rich thermodynamic behaviors and provides clues about the quantum nature of spacetime. The Hawking-Page phase transition between the Schwarzschild anti-de Sitter (AdS) black hole and the thermal AdS spacetime \cite{Page1983} provides a gravitational dual description of the confinement/deconfinement phase transition in the gauge field theory \cite{Witten1998}. The Reissner-Nordstr\"{o}m (RN) AdS black hole can undergo a first-order phase transition between small/large black holes, analogous to the liquid-gas phase transition in the Van der Waals fluid \cite{Chamblin1999a,Chamblin1999b}. In the extended phase space, where the cosmological constant is promoted to a thermodynamic variable and interpreted as a pressure \cite{Kastor2009}, a variety of thermodynamic phenomena have been investigated, including the $P-V$ criticality \cite{Kubiznak2012}, reentrant phase transitions \cite{Altamirano2013}, the heat engine \cite{Johson2014}, and superfluid-like criticality \cite{Hennigar2017}. A higher-dimensional origin of extended black hole thermodynamics has also been proposed through holographic braneworlds \cite{Frassino2023}. In addition, microscopic structures of black holes can be probed by the thermodynamic geometry \cite{Quevedo2009,Wei2015}.

The topological investigation provides a powerful approach to address universal and global properties of black hole solutions, without solving equations of motion and going into the details of solutions. Black hole solutions are identified as topological defects corresponding to zeros of a two-dimensional vector field \cite{SWWei2022,Hung2023,DiWu2023a,DiWu2023b,DiWu2023c,Chen2023,Chen2024,XDZhu2024,Zhang2024,Sajadi2025}, where the two-dimensional vector field is given by $\Vec{\phi}=(\partial_{r_+}\mathcal{F},-\cot{\Theta}\csc\Theta)$ with $\mathcal{F}$, $\Theta$, and $r_+$ being the off-shell Gibbs free energy \cite{York1986,RanLi2020,RanLi2022}, the auxiliary variable, and the event horizon radius, respectively. A negative (positive) winding number at each zero represents a locally unstable (stable) black hole. Summing the winding number of all zeros yields a topological charge that allows us to classify the thermodynamics of black holes into three classes \cite{SWWei2022}. The analysis of the asymptotic behavior of the Hawking temperatures also unveils new classes and subclasses \cite{WWei2024,Wu2025a,XDZhu2025,YChen2025,Wu2025b}.

The auxiliary variable $\Theta$ introduced in Ref. \cite{SWWei2022} is a mathematical tool for the purposes of the topological construction: its introduction amounts to an auxiliary completion of a one-dimensional problem because the particular $\phi$-mapping construction employed in Ref. \cite{SWWei2022} requires a two-component vector field for defining the relevant winding number. Thus, the auxiliary variable $\Theta$ is not a physical variable associated with an independent variation of the thermodynamic state, and therefore the $\Theta$-component of the two-dimensional vector field does not correspond to an additional equilibrium condition. This naturally raises the question of whether the thermodynamic topology can instead be formulated directly in terms of the thermodynamic state variables, without introducing an auxiliary variable solely to complete the vector field for the purpose of applying the $\phi$-mapping formalism.

The aim of the present work is to formulate the topological description of black hole thermodynamics in a manner more directly tied to the thermodynamic state space. More explicitly, we replace the vector field supplemented by the auxiliary $\Theta$-component with a vector field constructed directly from the gradient of the off-shell grand free energy with respect to the thermodynamic state variables $(S,Q,\ldots)$ associated with the ensemble parameters ($\tau_1,\tau_2,\ldots$). Thus, the vector field is defined entirely along the thermodynamic state-variable directions, with its components given by the derivatives of the off-shell grand free energy with respect to the corresponding thermodynamic state variables.

The use of the off-shell grand free energy in the present work should not be understood as a conceptual departure from the existing thermodynamic topology framework. Different thermodynamic ensembles have already been successfully explored by employing the thermodynamic potential appropriate to each ensemble \cite{Gogoi2023,Gogoi2024}. Rather, the essential point of the present construction is that the thermodynamic gradient itself supplies the multi-component vector field required for the topological study, without introducing the auxiliary variable. This construction naturally leads to a conserved higher-rank topological tensor and its topological flux which is given by the total topological index of the enclosed isolated zeros. These zeros are determined entirely within the thermodynamic state space and correspond to the stationary points of the off-shell grand free energy, and hence to the equilibrium black hole configurations. In this way, the topological construction is tied directly to the thermodynamic state space, establishing a more direct correspondence between the topological structure and the thermodynamic variables of the system.

In this framework, we define the asymptotic topological flux as the limiting value of the conserved topological flux through a family of finite hypersurfaces that are continuously pushed toward the asymptotic boundary of the ensemble parameter space. In this limit, the black hole geometry approaches its associated asymptotic/background spacetime geometry. It is therefore natural to expect that the asymptotic topological flux is determined by the asymptotic/background spacetime geometry, rather than by the detailed properties of the black hole, and is therefore invariant under continuous deformations that preserve this geometry. This implies that the asymptotic topological flux can serve as an invariant characterizing the asymptotic/background spacetime geometry, which is used to classify spacetime solutions into universal classes. Within this classification, we find that Minkowski/dS spacetimes are in the same class, whereas the AdS spacetime belongs to a distinct class. Interestingly, the flat-space limit of the AdS spacetime remains nontrivial in this classification. This observation may offer a complementary geometric viewpoint on the nontrivial behavior of the low-energy spectrum in the flat-space limit of the AdS spacetime suggested by the AdS distance conjecture \cite{Lust2019,Nam2023}. 

This paper is organized as follows. In Sec. \ref{simpl-case}, we construct a conserved rank-two topological tensor and its associated topological flux for black holes that can exchange both energy and matter with their environment, in the simple situation where they carry only electric charge. In Sec. \ref{generali}, we generalize this construction to arbitrary black hole solutions carrying multiple charges. In Sec. \ref{invariant}, we conjecture that the asymptotic topological flux serves as an invariant characterizing the asymptotic/background spacetime geometry and use it to classify spacetime solutions into universal classes. Finally, we present our conclusions in Sec. \ref{sec:conclu}. 

\section{\label{simpl-case}The simple case}

Let us consider the charged black hole, which is characterized by the entropy $S$ (or equivalently its event horizon radius $r_+$) and the electric charge $Q$. When the black hole can exchange both energy and matter with the environment due to its radiation and absorption, leading to a thermal and chemical equilibrium, the thermodynamic system is described by the off-shell grand free energy given by \cite{WBZhao2021,Hung2025}
\begin{eqnarray}
\hat{\mathcal{F}}=M-\frac{S}{\tau_1}-\frac{Q}{\tau_2},    
\end{eqnarray}
where $M$ is the Arnowitt-Deser-Misner (ADM) mass of the black hole, and $\tau_{1,2}$ are the inverse temperature and the inverse chemical potential of the ensemble, which are kept fixed. Additionally, the first law reads
\begin{eqnarray}
dM=TdS+\mu dQ,\label{frst-law-2D}     
\end{eqnarray}
where the Hawking temperature $T$ and the chemical potential $\mu$ are determined by
\begin{eqnarray}
T=\left(\frac{\partial M}{\partial S}\right)_{Q},\ \ \mu=\left(\frac{\partial M}{\partial Q}\right)_S.
\end{eqnarray}
 
We introduce a two-dimensional vector field $\Vec{\phi}$ defined as the gradient of the off-shell grand free energy $\hat{\mathcal{F}}$ with respect to the thermodynamic state variables $(S,Q)$, given by
\begin{eqnarray}
\Vec{\phi}=\nabla\hat{\mathcal{F}}=\left(\partial_S\hat{\mathcal{F}},\partial_Q\hat{\mathcal{F}}\right)\equiv(\phi^S,\phi^{Q}).   
\end{eqnarray}
The vector field $\Vec{\phi}$ determines the rate of change of $\hat{\mathcal{F}}$ with respect to the entropy $S$ and the electric charge $Q$. The vanishing of $\Vec{\phi}$ corresponds to the extreme points of $\hat{\mathcal{F}}$, which contribute dominantly to the partition function. On the other hand, the zeros of the vector field $\Vec{\phi}$, determined by $\tau_1=1/T$ and $\tau_2=1/\mu$, correspond to the black hole solutions. 

Including both the thermodynamic state variables $(S,Q)$ and the ensemble parameters $\tau_{1,2}$, we construct a four-dimensional space corresponding to the evolution of zeros of $\Vec{\phi}$ (topological defects in thermodynamic state space) when changing the ensemble parameters $\tau_1$ and $\tau_2$. In this space, we define a rank-two topological tensor as follows
\begin{eqnarray}
j^{\mu\nu}&=&\frac{1}{2\pi}\epsilon^{\mu\nu\rho\lambda}\epsilon_{ab}\partial_\rho n^a\partial_\lambda n^b,\label{simp-topo-tens}
\end{eqnarray}
where $\epsilon^{\mu\nu\rho\lambda}$ and $\epsilon_{ab}$ are the totally antisymmetric Levi-Civita tensors in four and two dimensions, respectively, $\partial_\rho\equiv(\partial_{\tau_1},\partial_{\tau_2},\partial_S,\partial_Q)$, and $n^a=\phi^a/||\Vec{\phi}||$ refers to the components of a normalized vector field $\Vec{n}=\Vec{\phi}/||\Vec{\phi}||$. The mathematical origin of the topological tensor $j^{\mu\nu}$ and its higher-dimensional generalization is represented in the appendix. We can decompose the topological tensor $j^{\mu\nu}$ into three components as $(j^{\tau_1\mu},j^{\tau_2\mu},j^{SQ})$, where $j^{\tau_1\mu}$ is the current flowing along the $\tau_2$-direction when adjusting the chemical potential of the ensemble, whereas, the current $j^{\tau_2\mu}$ flows along the $\tau_1$-direction when varying the ensemble temperature.

The rank-two (or higher-rank) topological tensor characterizes the topological defects associated with the zeros of the vector field $\Vec{\phi}$. Since $\Vec{\phi}$ is defined as the gradient of the off-shell grand free energy, its zeros correspond to stationary points of the off-shell grand free energy or, when the equilibrium conditions are satisfied, to black hole solutions. Its flux through a (hyper)surface at fixed ensemble parameters gives the total topological index of the isolated zeros contained within the (hyper)surface, as will be demonstrated below.

We compute the flux of the topological tensor through a chosen (two-dimensional) surface $\Sigma^{(2)}_{\{\tau_m\}}$ at fixed ensemble parameters $\{\tau_m\}$, which we refer to as the topological flux, as follows
\begin{eqnarray}
Q_{\{\tau_m\}}&=&\int_{\Sigma^{(2)}_{\{\tau_m\}}}j^{\tau_1\tau_2}d\Sigma_{\tau_1\tau_2}\nonumber\\
&=&\frac{1}{2\pi}\int_{\Sigma^{(2)}_{\{\tau_m\}}}\Delta_\phi\left(\ln||\Vec{\phi}||\right)J_2\left(\phi|x\right)d\Sigma_{\tau_1\tau_2}\label{Qt-int}\\
&=&\int_{\Sigma^{(2)}_{\{\tau_m\}}}\delta^2(\phi)J_2\left(\phi|x\right)d\Sigma_{\tau_1\tau_2},\nonumber
\end{eqnarray}
where $d\Sigma_{\tau_1\tau_2}=\frac{1}{2}\epsilon_{\tau_1\tau_2\mu\nu}dx^\mu\wedge dx^\nu$ is the surface element with $\wedge$ being the wedge product, $J_2\left(\phi|x\right)=\frac{1}{2}\epsilon^{\tau_1\tau_2\rho\lambda}\epsilon_{ab}\partial_\rho\phi^a\partial_\lambda\phi^b$ is the Jacobian of the transformation $x=(S,Q)\longrightarrow\phi=(\phi^S,\phi^{Q})$. Note that we have used the relation $\Delta_\phi\left(\ln||\Vec{\phi}||\right)=2\pi\delta^2(\phi)$ to obtain the third line in Eq. (\ref{Qt-int}). 

Although the components of the topological tensor depend on the specific choice of thermodynamic coordinates, they transform covariantly under the reparameterizations of the thermodynamic state space. Consequently, the quantity $Q_{\{\tau_m\}}$ is invariant under the reparameterizations that do not change the equilibrium configurations of the thermodynamic state space.

\begin{figure}[ht]
\begin{center}
\includegraphics[width=0.5\textwidth]{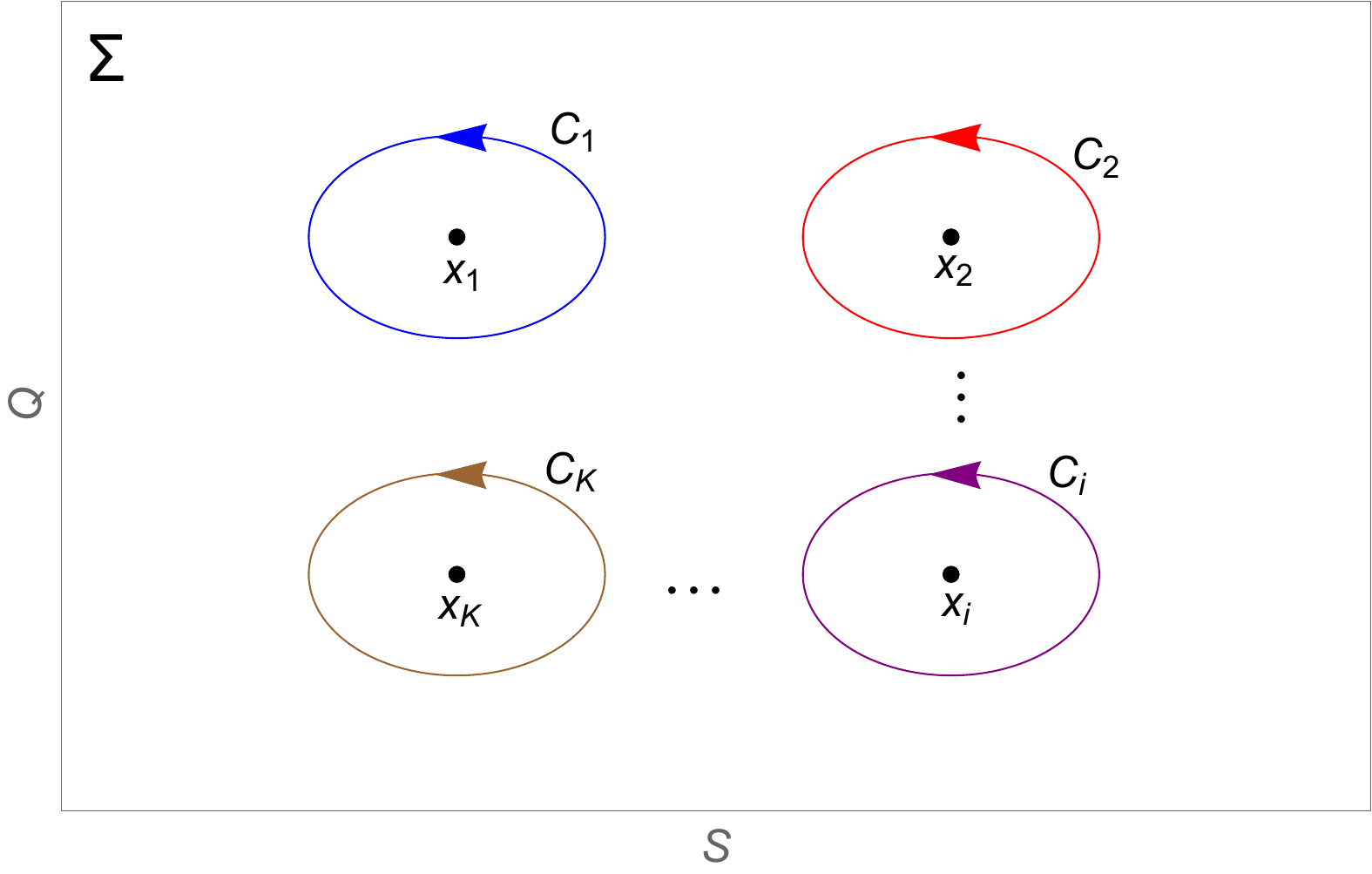}
\end{center}
\vspace{-0.5cm}
\caption{The thermodynamic state space $\Sigma$ at fixed ensemble parameters $(\tau_1,\tau_2)$ consists of $K$ topological defects corresponding to $K$ isolated zeros of the vector field $\Vec{\phi}$, denoted by $x_1$,..., $x_K$, which are marked by black dots, where each zero represents a black hole solution. The closed loops, denoted by $C_1$,..., $C_K$, are chosen such that each loop encloses one isolated zero of the vector field $\Vec{\phi}$.}\label{sigma}
\end{figure}

For all zeros contained within $\Sigma^{(2)}_{\{\tau_m\}}$ which are nondegenerate, i.e. $J_2\left(\phi|x\right)|_{x_i}\neq0$,\footnote{The Jacobian $J_2\left(\phi|x\right)|_{x_i}$ vanishes at second-order phase transition points. In this case, the bifurcation theory can be used to study the evolution of the topological currents \cite{YSDuan1998}.} we can express the delta function $\delta^2(\phi)$ as follows \cite{Schouten1951}
\begin{eqnarray}
\delta^2(\phi)=\sum_{x_i\in\Sigma^{(2)}_{\{\tau_m\}}}\frac{1}{|J_2\left(\phi|x\right)|_{x_i}|}\delta^2(x-x_i).\label{delta-2d}
\end{eqnarray}
By substituting this relation into Eq. (\ref{Qt-int}), we obtain $Q_{\{\tau_m\}}$ as follows
\begin{eqnarray}
Q_{\{\tau_m\}}=\sum_{x_i\in\Sigma^{(2)}_{\{\tau_m\}}}\text{index}_{x_i}(\Vec{\phi}),\label{topol-flux-ind}
\end{eqnarray}
where $\text{index}_{x_i}(\Vec{\phi})$ is the local topological index of the vector field $\Vec{\phi}$ at the zero $x_i$, which is computed by
\begin{eqnarray}
\text{index}_{x_i}(\Vec{\phi})&=&\text{sign}\left(J_2\left(\phi|x\right)|_{x_i}\right)\nonumber\\
&=&\frac{1}{2\pi}\int_{C_i}\epsilon_{ab}n^adn^b=\pm1,\label{index-comp}
\end{eqnarray}
where $C_i$ is a closed loop surrounding the isolated zero $x_i$, as depicted in Fig. \ref{sigma}. 
The quantity $\text{index}_{x_i}(\Vec{\phi})$ describes the behavior of the gradient of the off-shell grand free energy around its extreme point $x_i$. Thus, $\text{index}_{x_i}(\Vec{\phi})$ determines the thermodynamic properties of the corresponding black hole. More specifically, as seen later $\text{index}_{x_i}(\Vec{\phi})$ is related to the sign of the heat capacity of the black hole, hence the sign of $\text{index}_{x_i}(\Vec{\phi})$ would indicate the stability of the black hole solution corresponding to the zero $x_i$. 

The quantity $Q_{\{\tau_m\}}$ is formally defined as the flux of the topological tensor through the surface $\Sigma^{(2)}_{\{\tau_m\}}$. From Eq. (\ref{topol-flux-ind}), however, we find that the quantity $Q_{\{\tau_m\}}$ is equal to the total topological index of all isolated defects, corresponding to all equilibrium black hole configurations, contained within the surface at fixed ensemble parameters. Furthermore, since the topological tensor $j^{\mu\nu}$ is identically conserved, satisfying $\partial_\nu j^{\mu\nu}=0$, the quantity $Q_{\{\tau_m\}}$ is unchanged under a continuous deformation of the surface $\Sigma^{(2)}_{\{\tau_m\}}$, as long as the deformation does not cross any defect and no topological bifurcation occurs. Although the individual defects may change their locations under admissible continuous deformations, the sum of the local topological indices of the defects contained within the surface remains unchanged. Therefore, the quantity $Q_{\{\tau_m\}}$ captures global topological information that is insensitive to the detailed choice of the integration surface or to continuous variations of the ensemble parameters, provided that the topological structure of the space of equilibrium black hole configurations is preserved. In this sense, the quantity $Q_{\{\tau_m\}}$ is a global topological invariant of the thermodynamic configuration space. The same reasoning applies to the higher-dimensional generalization involving the rank-$N$ topological tensor and the corresponding $N$-dimensional hypersurface. 
\begin{figure*}
\begin{center}
\includegraphics[width=0.8\textwidth]{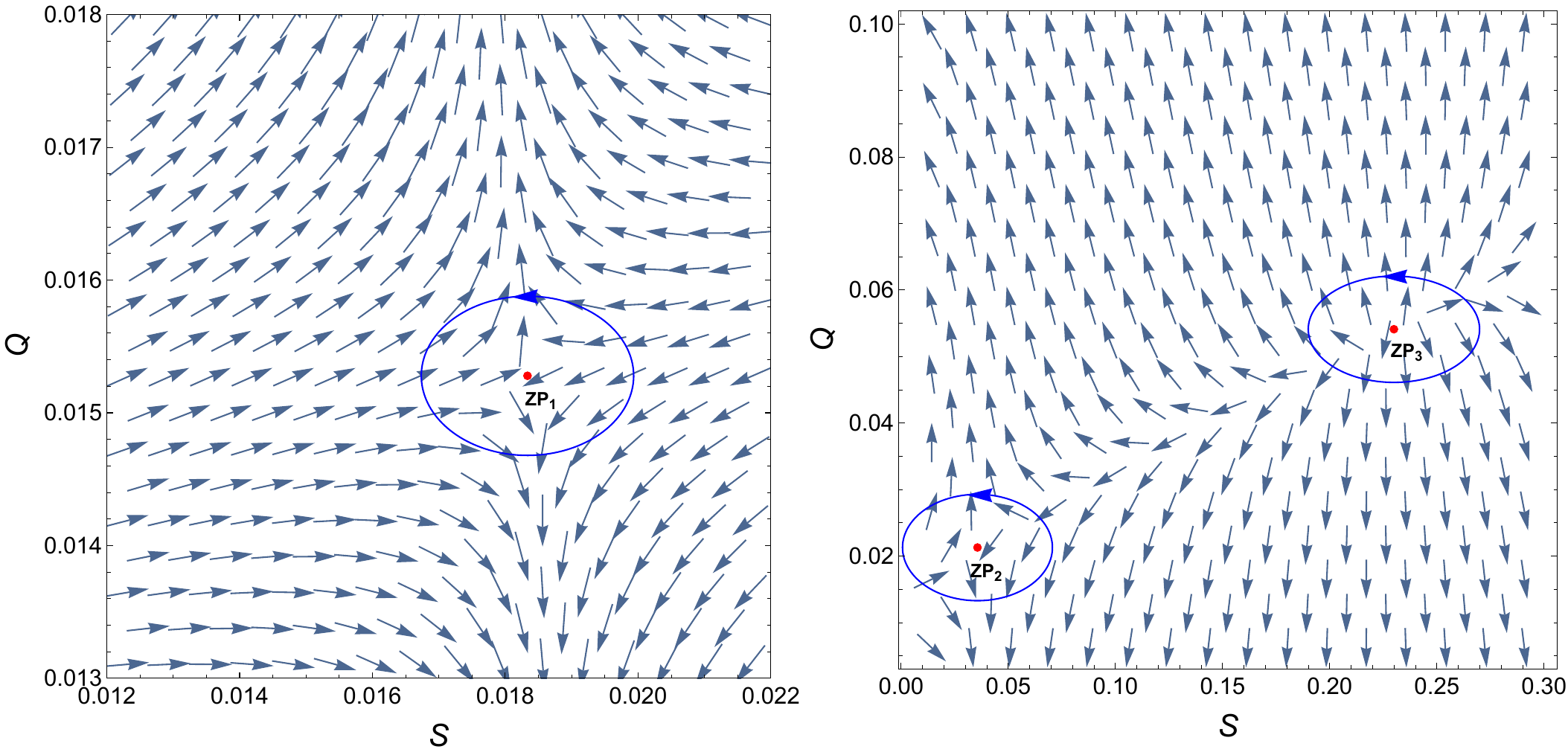}
\end{center}
\vspace{-0.5cm}
\caption{The left and right panels represent the unit vector field $\Vec{\phi}/||\Vec{\phi}||$ for the RN and RN-AdS black holes, respectively, with $\tau_1=1$ and $\tau_2=5$. We set the AdS curvature radius to $l=0.3$ for the RN-AdS black hole. Red dots refer to the zeros of $\Vec{\phi}$. The blue closed curves represent the contours $C_i$ in the thermodynamic state space, each enclosing one isolated zero, along which $\text{index}_{x_i}(\Vec{\phi})$ can be computed using Eq. (\ref{index-comp}).}\label{univect-RNBHs}
\end{figure*}

Different off-shell thermodynamic potentials generally have different gradient vector fields and, consequently, different stationary conditions. Their zero sets, which determine the locations and structure of the corresponding topological defects, are therefore not generally identical. As a result, the topological flux $Q_{\{\tau_m\}}$ is, in general, dependent on the choice of the thermodynamic potential and, more broadly, on the thermodynamic ensemble. The fluxes constructed from different off-shell thermodynamic potentials, such as the Helmholtz and Gibbs free energies, therefore characterize their respective topological defect structures and do not necessarily yield equivalent topological classifications. 

We find that the Jacobian $J_2\left(\phi|x\right)$, evaluated at $x_i$, can be expressed in terms of the heat capacity at the chemical potential kept fixed, $C_\mu=T\left(\partial S/\partial T\right)_\mu$, as follows
\begin{eqnarray}
J_2\left(\phi|x\right)|_{x_i}&=&\partial_ST\partial_{Q}\mu-\left(\partial_{Q}T\right)^2\nonumber\\
&=&\frac{T\partial_{Q}\mu}{C_\mu}.
\end{eqnarray}
This relation means that the sign of $J_2\left(\phi|x\right)|_{x_i}$ is the same as that of the heat capacity $C_\mu$ corresponding to the zero $x_i$. As a result, the locally stable black holes ($C_\mu>0$) would have $\text{index}_{x_i}(\Vec{\phi})=+1$. On the contrary, $\text{index}_{x_i}(\Vec{\phi})=-1$ describes the locally unstable black holes ($C_\mu<0$).

\begin{figure*}
\begin{center}
\includegraphics[width=0.6\textwidth]{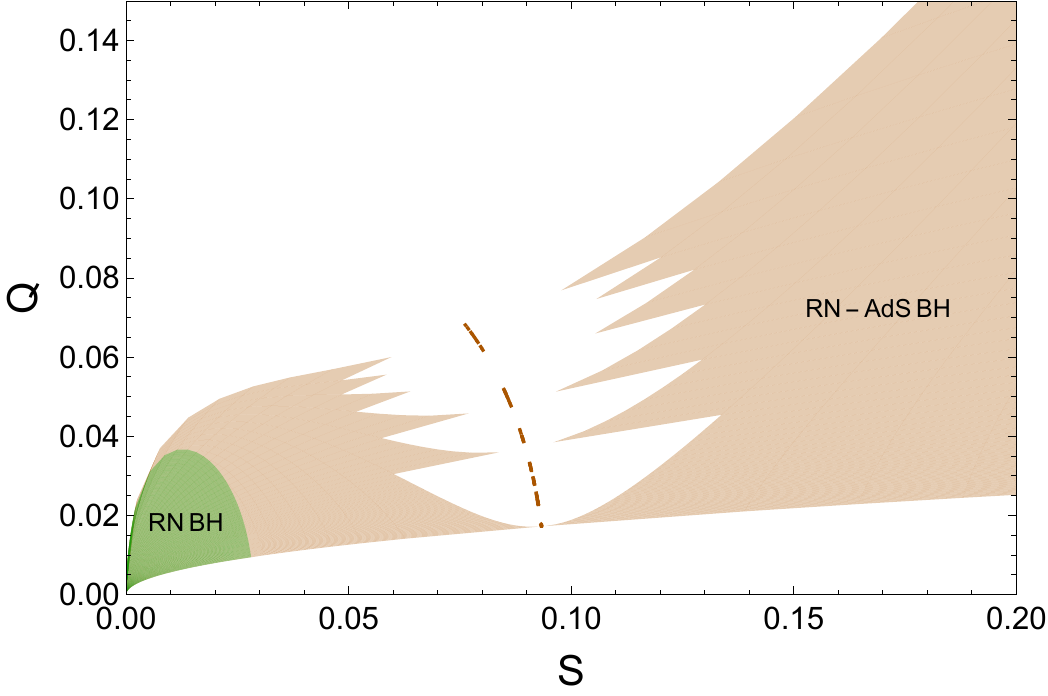}
\end{center}
\vspace{-0.5cm}
\caption{The dark green and dark orange regions refer to the zeros of the $\Vec{\phi}$ corresponding to the RN and RN-AdS black holes, respectively, for $\tau_1\in[0.01,1.2]$ and $\tau_2\in[1.01,10]$. For the RN-AdS black hole, we set the AdS curvature radius to $l=0.3$. The dashed curve with $\tau_1=2l\pi\tau_2/\sqrt{3(\tau^2_2-1)}$ corresponds to the generation points for the RN-AdS black hole.}\label{zero-plot-BHs}
\end{figure*}

For the RN black hole, which is the charged black hole in the Minkowski background, the corresponding normalized vector field $\Vec{n}$ is given by
\begin{eqnarray}
n^1&=&\left(\frac{S-\pi Q^2}{4\sqrt{\pi}S^{3/2}}-\frac{1}{\tau_1}\right)\left[\left(\frac{S-\pi Q^2}{4\sqrt{\pi}S^{3/2}}-\frac{1}{\tau_1}\right)^2+\left(Q\sqrt{\frac{\pi}{S}}-\frac{1}{\tau_2}\right)^2\right]^{-1/2},\\
n^2&=&\left(Q\sqrt{\frac{\pi}{S}}-\frac{1}{\tau_2}\right)\left[\left(\frac{S-\pi Q^2}{4\sqrt{\pi}S^{3/2}}-\frac{1}{\tau_1}\right)^2+\left(Q\sqrt{\frac{\pi}{S}}-\frac{1}{\tau_2}\right)^2\right]^{-1/2}.
\end{eqnarray}
The behavior of this unit vector field is depicted in the left panel of Fig. \ref{univect-RNBHs}, where it exhibits a zero denoted by the red dot ZP$_1$. One can indicate that the vector field $\Vec{\phi}$ corresponding to the RN black hole possesses a unique zero determined by 
\begin{eqnarray}
\sqrt{S}&=&\frac{\tau_1(\tau^2_2-1)}{4\sqrt{\pi}\tau^2_2},\\
Q&=&\frac{1}{\tau_2}\sqrt{\frac{S}{\pi}}.
\end{eqnarray}
Explicitly, we show the parameter region for the zero of the $\Vec{\phi}$ in the $(S,Q)$ plane with $\tau_1\in[0.01,1.2]$ and $\tau_2\in[1.01,10]$, which is given by the dark green region in Fig. \ref{zero-plot-BHs}.
This zero has $\text{index}_{x_i}(\Vec{\phi})=-1$ because the heat capacity at the fixed chemical potential, $C_\mu=-2S$, is always negative. As a result, the RN black hole has the topological flux $Q_{\{\tau_m\}}=-1$.

For the RN-AdS black hole, which is the charged black hole in the AdS background, its off-shell grand free energy is expressed as follows \cite{SBanerjee2010}
\begin{eqnarray}
\hat{\mathcal{F}}=\frac{1}{2}\sqrt{\frac{S}{\pi}}\left(1+\frac{\pi Q^2}{S}+\frac{S}{\pi l^2}\right)-\frac{S}{\tau_1}-\frac{Q}{\tau_2},\label{RN-AdS-Freeenrg}   
\end{eqnarray}
where $l$ is the curvature radius of the asymptotic AdS spacetime. In the present construction, the cosmological constant $\Lambda=-3/l^2$ is treated as a fixed background parameter. Hence, the cosmological constant does not belong to the thermodynamic state space on which the gradient of the off-shell grand free energy is taken. Even the cosmological constant is promoted to a thermodynamic variable, which is the thermodynamic pressure $P=-\Lambda/(8\pi)$, leading to the extended phase space as studied in Ref. \cite{SWWei2022}, the present construction can be derived from the framework of extended black hole thermodynamics with $P$ kept fixed if the ensemble does not change the thermodynamic volume $V=(\partial M/\partial P)_{S,Q}$.

\begin{figure}[ht]
  \includegraphics[scale=0.5]{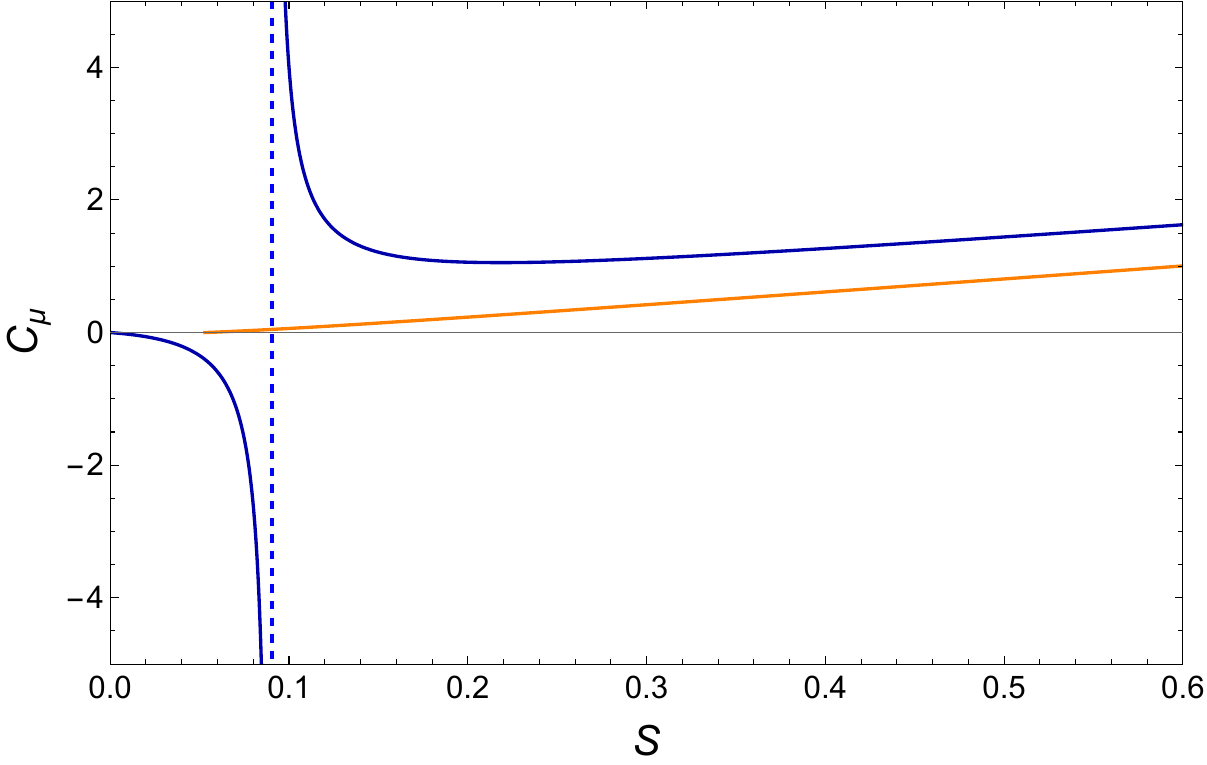}
  \caption{The behavior of the RN-AdS black hole heat capacity at the chemical potential kept fixed in the entropy $S$ with the AdS radius $l=0.3$. The blue and orange curves correspond to $\tau_2=5$ and $\tau_2=0.8$, respectively.}
  \label{temp-RNAdSBH}
\end{figure}

From the expression of the off-shell grand free energy, given in Eq. (\ref{RN-AdS-Freeenrg}), we get the normalized vector field $\Vec{n}$ associated with the RN-AdS black hole as
\begin{eqnarray}
n^1&=&\left[\frac{3S^2+l^2\pi\left(S-\pi Q^2\right)}{4l^2(S\pi)^{3/2}}-\frac{1}{\tau_1}\right]\left\{\left[\frac{3S^2+l^2\pi\left(S-\pi Q^2\right)}{4l^2(S\pi)^{3/2}}-\frac{1}{\tau_1}\right]^2+\left(Q\sqrt{\frac{\pi}{S}}-\frac{1}{\tau_2}\right)^2\right\}^{-\frac{1}{2}},\\
n^2&=&\left(Q\sqrt{\frac{\pi}{S}}-\frac{1}{\tau_2}\right)\left\{\left[\frac{3S^2+l^2\pi\left(S-\pi Q^2\right)}{4l^2(S\pi)^{3/2}}-\frac{1}{\tau_1}\right]^2+\left(Q\sqrt{\frac{\pi}{S}}-\frac{1}{\tau_2}\right)^2\right\}^{-\frac{1}{2}}.
\end{eqnarray}
If the inverse chemical potential of the ensemble is kept fixed in the region of $\tau_2>1$ (or $\mu<1$), the vector field $\Vec{\phi}$ has two distinguished zeros, which are determined by
\begin{eqnarray}
\sqrt{S}&=&\frac{2l^2\pi^{3/2}}{3\tau_1}\left[1\pm\sqrt{1-\frac{3\tau^2_1(\tau^2_2-1)}{4l^2\pi^2\tau^2_2}}\right],\label{ent-sol-RNAdS}\\
Q&=&\frac{1}{\tau_2}\sqrt{\frac{S}{\pi}},
\end{eqnarray}
where the solutions with the signs ``$-$" and ``+" are denoted by ZP$_2$ and ZP$_3$ as shown in the right panel of Fig. \ref{univect-RNBHs}. The space of these zeros in the region of $\tau_1\in[0.01,1.2]$ and $\tau_2\in[1.01,10]$, with $l=0.3$, is depicted in Fig. \ref{zero-plot-BHs}, corresponding to the dark orange region. With the heat capacity at the chemical potential kept fixed, given by 
\begin{eqnarray}
C_\mu=-2S\left[1-\frac{6S}{3S-\pi l^2(1-\mu^2)}\right],    
\end{eqnarray}
we can indicate that the ZP$_2$ and ZP$_3$ zeros have $\text{index}_{x_i}(\Vec{\phi})=-1$ and $+1$, corresponding to the locally unstable and stable black hole solutions. This means that the topological flux of the RN-AdS black hole in the case where we fix the inverse chemical potential in the region of $\tau_2>1$ is given by $Q_{\{\tau_m\}}=-1+1=0$. 
In addition, the heat capacity $C_\mu$ in the region of $\tau_2>1$, which corresponds to the blue curve given in Fig. \ref{temp-RNAdSBH}, exhibits a divergence at $S=\pi l^2(1-\mu^2)/3$, implying a second-order phase transition between the small and large black holes. 
However, if we fix the inverse chemical potential in the region of $\tau_2<1$ (or $\mu>1$), there exists only one zero which is the solution with the sign ``+" given in Eq. (\ref{ent-sol-RNAdS}), corresponding to $\text{index}_{x_i}(\Vec{\phi})=+1$. Hence, in this case, the topological flux of the RN-AdS black hole is $Q_{\{\tau_m\}}=1$. We observe from the orange curve given in Fig. \ref{temp-RNAdSBH} that the heat capacity behaves continuously, thus there is no phase transition. With the combined regions of the ensemble parameter $\tau_2$, we find the topological flux of the RN-AdS black hole as $Q_{\{\tau_m\}}\in\{0,1\}$.

Analogously, with the RN-dS black hole, which is asymptotic to the dS geometry, its vector field possesses a unique zero, which is given by
\begin{eqnarray}
\sqrt{S}&=&\frac{2l^2\pi^{3/2}}{3\tau_1}\left[\sqrt{1+\frac{3\tau^2_1(\tau^2_2-1)}{4l^2\pi^2\tau^2_2}}-1\right],\label{ent-sol-RNdS}\\
Q&=&\frac{1}{\tau_2}\sqrt{\frac{S}{\pi}},   
\end{eqnarray}
where $l$ refers to the curvature radius of the dS background space. With the corresponding heat capacity $C_\mu=-2S\left[1-6S/\left(3S+\pi l^2(1-\mu^2)\right)\right]$, we can show that this zero corresponds to the negative heat capacity. As a result, the topological flux of the RN-dS black hole is $Q_{\{\tau_m\}}=-1$, which is analogous to the RN black hole.

It is important to emphasize that, throughout the above analysis and the entire paper, the cosmological constant, or equivalently the AdS radius, is kept fixed. A reason for keeping the cosmological constant fixed in the present construction is that we seek to interpret the asymptotic topological flux in the ensemble parameter space as an invariant associated with the asymptotic/background spacetime geometry, as analyzed below. While varying the cosmological constant changes the asymptotic AdS radius, it changes the asymptotic/background spacetime geometry itself. However, the extended phase space contains fixed-$P$ hypersurfaces on which the asymptotic/background geometry is unchanged; the present construction is precisely the restriction of the extended construction to such a hypersurface.

When the cosmological constant is treated as the thermodynamic pressure, the gradient space is enlarged from $(S,Q)$ to $(S,Q,P)$, which also enlarges the ensemble parameter space. As a result, the normalized vector field $\Vec{n}$ acquires an additional $P$-component, which defines an $S^2$-valued map, rather than the $S^1$-valued map corresponding to the case of the cosmological constant kept fixed. This leads to a higher-dimensional topological tensor associated with a different topological flux, which is relevant to the degree of the $S^2$-valued map. The fully extended $(S,Q,P)$ construction is an interesting generalization, which is beyond the scope of the present work and will be considered in our future work.

\section{\label{generali} Generalization}

We generalize the topological study of the black hole thermodynamics, represented above, to black holes which are characterized by the entropy $S$ and other quantities denoted by $\{\Tilde{Q}_2,\Tilde{Q}_3,...,\Tilde{Q}_{N}\}$. The corresponding first law reads
\begin{eqnarray}
dM=TdS+\sum^{N}_{m=2}\mu_md\Tilde{Q}_m,    
\end{eqnarray}
where the AMD mass $M$ is a function of the entropy $S$ and the quantities $\{\Tilde{Q}_2,\Tilde{Q}_3,...,\Tilde{Q}_{N}\}$, and $\mu_m$ is the conjugate to $\Tilde{Q}_m$ which is called the generalized chemical potential. The Hawking temperature and the generalized chemical potentials are determined by
\begin{eqnarray}
 T&=&\left(\frac{\partial M}{\partial S}\right)_{\Tilde{Q}_2,..,\Tilde{Q}_{N}},\\
\mu_m&=&\left(\frac{\partial M}{\partial\Tilde{Q}_m}\right)_{S,\Tilde{Q}_2,..,\Tilde{Q}_{m-1},\Tilde{Q}_{m+1},...,\Tilde{Q}_{N}}.   
\end{eqnarray}

When the black hole is in contact with the environment where it can exchange both energy and matter, the black hole thermodynamics is described by the following off-shell grand free energy
\begin{eqnarray}
\hat{\mathcal{F}}=M-\frac{S}{\tau_1}-\sum^{N}_{m=2}\frac{\Tilde{Q}_m}{\tau_m},    
\end{eqnarray}
where $\{\tau_m$\} with $m=2,..., N$ are the inverse generalized chemical potentials of the ensemble.

The gradient of the off-shell grand free energy with respect to $N$ thermodynamic state variables $(S,\Tilde{Q}_2,...,\Tilde{Q}_{N})$ defines the vector field $\Vec{\phi}=\nabla\hat{\mathcal{F}}$, given by
\begin{eqnarray}
\Vec{\phi}=\left(\partial_S\hat{\mathcal{F}},\partial_{\Tilde{Q}_2}\hat{\mathcal{F}},...,\partial_{\Tilde{Q}_N}\hat{\mathcal{F}}\right)\equiv\left(\phi^S,\phi^{\Tilde{Q}_2},...,\phi^{\Tilde{Q}_{N}}\right).   
\end{eqnarray}
The zeros of this vector field are determined by $T=1/\tau_1$ and $\mu_m=1/\tau_m$, with $m=2,..., N$, corresponding to black hole solutions. We introduce a totally antisymmetric rank-$N$ topological tensor as follows
\begin{eqnarray}
j^{\nu_1...\nu_N}&=&\frac{1}{\omega_{N-1}}\epsilon^{\nu_1...\nu_N\rho_1...\rho_N}\epsilon_{a_1...a_N}\partial_{\rho_1}n^{a_1}...\partial_{\rho_N} n^{a_N},\label{gen-topo-tens}
\end{eqnarray}
where $\omega_{N-1}=2\pi^{N/2}/\Gamma(\frac{N}{2})$ is the surface area of the $(N-1)$-dimensional unit sphere. This topological tensor consists of $N$ currents which flow along the ensemble-parameter directions $\tau_1$, $\tau_2$,...,$\tau_{N}$ corresponding to the change of the inverse temperature and the inverse generalized chemical potentials of the ensemble. 

The topological current introduced in Ref. \cite{SWWei2022} is constructed from the normalized two-component vector field, which defines a map from the three-dimensional space specified by $(r_h,\Theta,\tau)$ to the unit circle $S^1$. Its associated topological charge is the winding number of this map around the zeros of the associated two-dimensional vector field. The higher-rank topological tensor $j^{\nu_1...\nu_N}$ introduced in the present work is the corresponding higher-dimensional generalization of the $\phi$-mapping construction, corresponding to the following map $\Vec{n}: \mathcal{M}^{(2N)}\setminus Z(\phi)\longrightarrow S^{N-1}$, where $\mathcal{M}^{(2N)}$ denotes the $2N$-dimensional space of thermodynamic state variables and ensemble parameters, $Z(\phi)$ denotes the zero set of the higher-dimensional vector field $\Vec{\phi}$, and $S^{N-1}$ is the $(N-1)$-dimensional sphere. Such a higher-rank topological tensor has been developed in the $\phi$-mapping framework for $\Tilde{p}$-brane defects associated with magnetic sources \cite{Duan2000}.

The flux of the topological tensor $j^{\nu_1...\nu_N}$ through a $N$-dimensional hypersurface $\Sigma^{(N)}_{\{\tau_m\}}$ at fixed ensemble parameters $\{\tau_m\}$ reads
\begin{eqnarray}
Q_{\{\tau_m\}}&=&\int_{\Sigma^{(N)}_{\{\tau_m\}}}j^{\tau_1...\tau_N}d\Sigma_{\tau_1\ldots\tau_N}\nonumber\\
&=&-\frac{1}{\omega_{N-1}}\int_{\Sigma^{(N)}_{\{\tau_m\}}}\Delta_\phi\left(\frac{1}{||\Vec{\phi}||^{N-2}}\right)J_{N}\left(\phi|x\right)d\Sigma_{\tau_1\ldots\tau_N},
\label{HD-TC}
\end{eqnarray}
where $d\Sigma_{\tau_1\ldots\tau_N}$ is an $N$-dimensional surface element written as
\begin{equation}
d\Sigma_{\tau_1\ldots\tau_N}=\frac{1}{N!}\epsilon_{\tau_1\ldots\tau_N\nu_1\ldots\nu_N}
dx^{\nu_1}\wedge\ldots\wedge dx^{\nu_N},
\end{equation}
and $J_{N}\left(\phi|x\right)$ is the Jacobian of the transformation $x=(S,\Tilde{Q}_2,...,\Tilde{Q}_N)\longrightarrow\phi=(\phi^S,\phi^{\Tilde{Q}_2},...,\phi^{\Tilde{Q}_N})$, given by
\begin{eqnarray}
J_{N}\left(\phi|x\right)&=&\frac{1}{N!}\epsilon^{\tau_1...\tau_N\nu_1...\nu_N}\epsilon_{a_1...a_N}\partial_{\nu_1}\phi^{a_1}...\partial_{\nu_N}\phi^{a_N}\nonumber\\   
&=&\begin{vmatrix}
\partial_{\Tilde{Q}_2}\mu_2 & \ldots & \partial_{\Tilde{Q}_2}\mu_N\\
\vdots & \ldots & \vdots\\
\partial_{\Tilde{Q}_N}\mu_2 & \ldots & \partial_{\Tilde{Q}_N}\mu_N
\end{vmatrix}\frac{T}{C_{\mu_2,...,\mu_N}},\label{Jacob-HD}
\end{eqnarray}
where $|...|$ denotes the determinant of the $(N-1)\times(N-1)$ matrix $\partial_{\Tilde{Q}_m}\mu_n$ and $C_{\mu_2,...,\mu_N}=T\left(\partial S/\partial T\right)_{\mu_2,...,\mu_N}$ is the heat capacity at the generalized chemical potentials kept fixed. By substituting the following relations 
\begin{eqnarray}
\Delta_\phi\left(\frac{1}{||\Vec{\phi}||^{N-2}}\right)&=&-\omega_{N-1}\delta^{N}\big(\phi\big),\\
\delta^{N}(\phi)&=&\sum_{x_i\in\Sigma^{(N)}_{\{\tau_m\}}}\frac{\delta^{N}(x-x_i)}{|J_{N}\left(\phi|x\right)|_{x_i}|},\label{delta-Nd}   
\end{eqnarray}
into the second line of Eq. (\ref{HD-TC}), we compute the topological flux $Q_{\{\tau_m\}}$ as  
\begin{eqnarray}
Q_{\{\tau_m\}}&=&\sum_{x_i\in\Sigma^{(N)}_{\{\tau_m\}}}\text{index}_{x_i}(\Vec{\phi})\nonumber\\
&=&\sum_{x_i\in\Sigma^{(N)}_{\{\tau_m\}}}\text{sign}\left(J_{N}\left(\phi|x\right)|_{x_i}\right).
\end{eqnarray}
If the determinant given in Eq. (\ref{Jacob-HD}) is positive, then $\text{index}_{x_i}(\Vec{\phi})$ is determined by the sign of the heat capacity $C_{\mu_2,...,\mu_N}$. This means that the black hole corresponding to the zero $x_i$ would be locally stable if $\text{index}_{x_i}(\Vec{\phi})=+1$, whereas $\text{index}_{x_i}(\Vec{\phi})=-1$ indicates the locally unstable black hole. On the contrary, if the determinant is negative, the $\text{index}_{x_i}(\Vec{\phi})=+1$ and $-1$ correspond to the black holes which are locally unstable and stable, respectively.

Due to the $\phi$-mapping identities, given in Eqs. (\ref{delta-2d}) and (\ref{delta-Nd}), for isolated nondegenerate zeros, the flux $Q_{\{\tau_m\}}$ is expressed as the sum of the local topological indices of the vector field $\Vec{\phi}$. In the two-dimensional case, the local topological index reduces to the local winding number, so that $Q_{\{\tau_m\}}$ reproduces the same topological invariant as the winding-number construction of Ref. \cite{SWWei2022}. In higher dimensions, the local topological index is the degree of the corresponding map $S^{N-1}\longrightarrow S^{N-1}$, and the flux $Q_{\{\tau_m\}}$ gives the total topological index of the enclosed zeros of the vector field $\Vec{\phi}$. Thus, the topological status of $Q_{\{\tau_m\}}$ follows from its identification with an integer-valued index, rather than merely from the conservation of the higher-rank tensor.

We can check that the rank-$N$ topological tensor $j^{\nu_1...\nu_N}$ is identically conserved, satisfying
\begin{eqnarray}
\partial_{\nu_k}j^{\nu_1...\nu_k...\nu_N}=0,\ \ \text{with}\ \ k=1,...,N.\label{conser-gen-tpten}   
\end{eqnarray}
This local conservation law implies that $Q_{\{\tau_m\}}$ is invariant under an admissible continuous deformation of the integration hypersurface $\Sigma^{(N)}_{\{\tau_m\}}$, as long as the deformation does not cross the zero set and no relevant boundary contribution appears. Consequently, for a continuous family of integration hypersurfaces parametrized by the ensemble parameters, the topological flux $Q_{\{\tau_m\}}$ satisfies
\begin{eqnarray}
\frac{\partial Q_{\{\tau_m\}}}{\partial\tau_k}=0 \ \ \text{with}\ \ k=1,...,N.\label{top-char-HD-conser}
\end{eqnarray}
More explicitly, this ensemble-parameter independence requires: (i) the integration hypersurfaces can be continuously deformed into one another without crossing the zero set; (ii) the relevant boundary contribution at infinity vanishes; (iii) the enclosed zeros remain isolated and nondegenerate, so that their local topological indices are well defined. Under these assumptions, if $\Sigma^{(N)}_{\{\tau_m\}}$ and $\Sigma^{(N)}_{\{\tau'_m\}}$ are two continuously deformable integration hypersurfaces corresponding to different values of the ensemble parameters, then
\begin{eqnarray}
Q_{\{\tau_m\}}=Q_{\{\tau'_m\}}.   
\end{eqnarray}

\begin{figure}[ht]
  \includegraphics[scale=0.5]{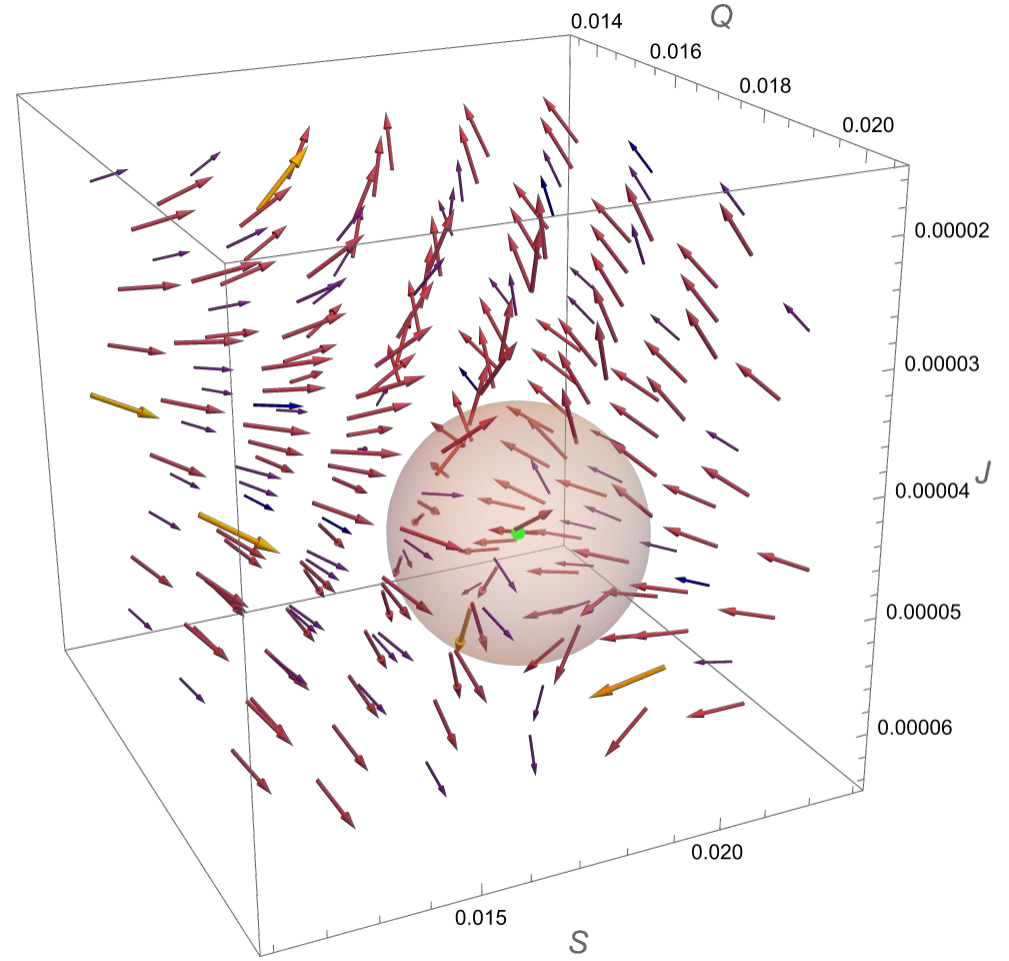}
  \caption{The behavior of $\Vec{\phi}/||\Vec{\phi}||$ in the $(S,Q,J)$ space with $\tau_1=1$, $\tau_2=4$, and $\tau_3=5$. The green dot refers to a zero of $\Vec{\phi}$. The light orange closed surface refers to the surface $\mathcal{S}_i$, enclosing the zero, by which we compute $\text{index}_{x_i}(\Vec{\phi})=\frac{1}{\omega_2}\int_{\mathcal{S}_i}\epsilon_{abc}n^{a}dn^{b}\wedge dn^{c}$.}
  \label{KNBH-vf}
\end{figure}

Based on the framework developed above, let us study the thermodynamic topology of the Kerr-Newman black hole \cite{Kerr1963,Newman1965} with the charges $\{\Tilde{Q}_2,\Tilde{Q}_3\}$ corresponding to the electric charge $Q$ and the angular momentum $J$. Its off-shell grand free energy reads
\begin{eqnarray}
\hat{\mathcal{F}}=M-\frac{S}{\tau_1}-\frac{Q}{\tau_2}-\frac{J}{\tau_3}.    
\end{eqnarray}
The corresponding vector field is
\begin{eqnarray}
\Vec{\phi}=\nabla\hat{\mathcal{F}}=\left(T-\frac{1}{\tau_1},\mu-\frac{1}{\tau_2},\Omega-\frac{1}{\tau_3}\right),    
\end{eqnarray}
where the Hawking temperature $T$, the chemical potential $\mu$, and the angular velocity $\Omega$ are given by 
\begin{eqnarray}
T&=&\frac{r_+}{4\pi(r^2_++a^2)}\left(1-\frac{a^2}{r^2_+}-\frac{Q^2}{r^2_+}\right),\\
\mu&=&\frac{Qr_+}{r^2_++a^2},\\
\Omega&=&\frac{a}{r^2_++a^2},
\end{eqnarray}
where the parameter $a$ and the horizon radius $r_+$ are related to $S$ and $J$ as 
\begin{eqnarray}
a&=&2J\sqrt{\frac{\pi S}{4J^2\pi^2+(S+\pi Q^2)^2}},\\    
r_+&=&\sqrt{\frac{S}{\pi}-a^2}.
\end{eqnarray}
The behavior of $\Vec{\phi}/||\Vec{\phi}||$ is depicted in Fig. \ref{KNBH-vf} with the ensemble parameters given by $\tau_1=1$, $\tau_2=4$, and $\tau_3=5$.
A zero of $\Vec{\phi}$ is located at $S\approx0.01747$, $Q\approx0.01865$, and $J\approx0.00004$. The black hole solution corresponding to this zero has a negative heat capacity $C_{\mu,\Omega}$ (the heat capacity at the chemical potential and angular velocity kept fixed), which indicates its thermodynamic instability. In addition, we show that this black hole solution has $\text{index}_{x_i}(\Vec{\phi})=-1$, leading to the topological flux $Q_{\{\tau_m\}}=-1$.

The Hawking temperature of the Kerr-Newman black hole, with the chemical potential and angular velocity kept fixed, reads
\begin{eqnarray}
T=\frac{\pi^2(1-\mu^2)-S\Omega^2\left(3\pi-2S\Omega^2\right)}{4\pi\sqrt{S}\left(\pi-S\Omega^2\right)^{3/2}},    
\end{eqnarray}
with $\mu^2<1$ and $S\Omega^2<\pi$ for the existence of the black hole solution. We observe that the Hawking temperature behaves continuously when varying the ensemble parameters $\tau_2=1/\mu$ and $\tau_3=1/\Omega$. This means that the topological flux $Q_{\{\tau_m\}}$ for the Kerr-Newman black hole is always equal to $-1$. 

The thermodynamic topology studies have already been extended to $f(R)$ gravity \cite{Hazarika2024}, Einstein-Gauss-Bonnet gravity \cite{Shahzad2023}, massive gravity \cite{Chen2023}, and other modified gravity models \cite{NCBai2023,DChen2024,BHazarika2024,Gashti2025}. Thus, the proposed construction is not restricted to the black hole thermodynamics in the Einstein-Maxwell theory. The essential ingredients of the present formalism are the existence of an appropriate off-shell thermodynamic potential and a thermodynamic state space containing the relevant thermodynamic variables corresponding to the ensemble variables. Besides the entropy, electric or magnetic charges, and angular momentum, when the additional scalar charges or additional gravity couplings constitute independent thermodynamic degrees of freedom, they can be incorporated into the thermodynamic state space. Consequently, higher-curvature gravities, scalarized black holes, and other modified gravity theories provide natural settings in which to apply the present framework.

\section{\label{invariant} Asymptotic topological flux}

The asymptotic topological flux is obtained as the limit of the topological flux $Q_{\{\tau_m\}}$ through a continuous family of finite hypersurfaces, given by 
\begin{eqnarray}
Q_{\infty}=\lim_{\{\tau_m\rightarrow\infty\}}\int_{\Sigma^{(N)}_{\{\tau_m\}}}j^{\tau_1...\tau_N}d\Sigma_{\tau_1\ldots\tau_N}.   
\end{eqnarray}
It is important to emphasize that $Q_{\infty}$ should not be interpreted as the net topological index of all isolated zeros located at infinity. Rather, it represents the asymptotic value of the topological flux carried by the defects contained within the finite hypersurfaces as these hypersurfaces are continuously deformed toward the asymptotic boundary of the ensemble parameter space. Thus, the topological information associated with $Q_{\infty}$ is encoded in the entire continuous family of finite hypersurfaces approaching the asymptotic boundary. 

If no additional zero enters or leaves the hypersurface as it is deformed toward infinity, no additional boundary contribution is generated, and the Jacobian $J_{N}\left(\phi|x\right)\neq0$ at all zeros encountered along the deformation, then the asymptotic topological flux is equal to the topological flux through any admissible finite hypersurface belonging to the same deformation class:
\begin{eqnarray}
Q_{\infty}=Q_{\text{finite}}.\label{asymp-tflux-com}    
\end{eqnarray}
Consequently, in order to determine $Q_{\infty}$ from the relation (\ref{asymp-tflux-com}), the finite hypersurface must belong to the same connected admissible ensemble-parameter sector as the asymptotic limit $\{\tau_m\rightarrow\infty\}$. A hypersurface lying in a region that cannot be continuously connected to the asymptotic limit without crossing a critical or singular region, or otherwise changing the enclosed defect structure, cannot be used to establish the asymptotic topological flux. For the RN-AdS black hole, for example, the region $\tau_2<1$ is separated from the asymptotic sector $\tau_2>1$ in this sense. Using the relation (\ref{asymp-tflux-com}), we find the asymptotic topological flux associated with the RN, RN-dS, and Kerr-Newman black holes as $Q_{\infty}=-1$. With the RN-AdS black hole, $Q_{\infty}$ is linked to $Q_{\text{finite}}$ only in the region of $\tau_2>1$. Thus, the asymptotic topological flux associated with the RN-AdS black hole is $Q_{\infty}=0$.

In the limit $(\tau_1,\tau_2,...,\tau_N)\rightarrow\infty$, the corresponding thermodynamic state variables satisfy $(S,\Tilde{Q}_2,...,\Tilde{Q}_{N})\rightarrow0$. Consequently, the black hole geometry approaches the associated asymptotic/background spacetime geometry:
\begin{eqnarray}
(\tau_1,\tau_2,...,\tau_N)\rightarrow\infty \ \ \implies \ \ (S,\Tilde{Q}_2,\Tilde{Q}_3,...,\Tilde{Q}_{N})\rightarrow0 \ \ \implies \ \ g^{\text{BH}}_{\mu\nu}\rightarrow g^{\text{background}}_{\mu\nu},    
\end{eqnarray}
where $g^{\text{BH}}_{\mu\nu}$ and $g^{\text{background}}_{\mu\nu}$ denote the black hole metric and the corresponding asymptotic/background spacetime metric, respectively. This observation motivates the interpretation of the asymptotic topological flux as a quantity determined by the asymptotic/background spacetime geometry, namely,
\begin{eqnarray}
Q_{\infty}=Q_{\infty}[g^{\text{background}}_{\mu\nu}].\label{Qt-asym}    
\end{eqnarray}
If the asymptotic topological flux depends only on the asymptotic/background spacetime geometry, then any admissible continuous deformation that preserves the relevant asymptotic/background geometry must leave $Q_{\infty}$ unchanged. Therefore, we conjecture that $Q_{\infty}$ defines an invariant associated with the asymptotic/background spacetime geometry. For example, the RN and Kerr-Newman black holes are distinct spacetime solutions represent distinct spacetime solutions, but both approach the same asymptotic Minkowski geometry. Correspondingly, they yield the same asymptotic topological flux, $Q_{\infty}=-1$, suggesting that $Q_{\infty}$ is determined by the common asymptotic/background geometry rather than by the specific black hole solution.

Based on the asymptotic topological flux $Q_\infty$, we find that the maximally symmetric constant-curvature spacetimes considered here fall into two classes, as summarized in 
Table. \ref{table1}.
\begin{table}[h]
\setlength{\tabcolsep}{3mm}{
\begin{center}
\begin{tabular}{cc}
  \hline\hline
 Kind of geometries & \ \ \ \ $Q_{\infty}$\ \ \ \ \\\hline
Minkowski/dS spacetimes & -1\\ 
AdS spacetime & 0\\
\hline\hline
\end{tabular}
\caption{The universal classes of the maximally symmetric constant-curvature spacetimes based on the asymptotic topological flux $Q_\infty$.}\label{table1}
\end{center}}
\end{table}
Minkowski and dS spacetimes have the same value, $Q_\infty=-1$, whereas AdS spacetime has $Q_\infty=0$. Thus, within the class of geometries examined in this work, the asymptotic topological flux distinguishes the AdS geometry from the Minkowski/dS geometries.

It is interesting that this classification has a geometric counterpart in the embedding properties of the corresponding spacetimes. Minkowski and dS spacetimes can be realized within a higher-dimensional flat embedding space with a single time-like direction, whereas an embedding of AdS spacetime requires a higher-dimensional flat space with two time-like directions \cite{Natsuume2014}. In this sense, the distinction indicated by $Q_{\infty}$ is consistent with a known qualitative distinction between the embedding structures of AdS and Minkowski/dS spacetimes.

The fact that Minkowski and dS spacetimes belong to the same class, characterized by $Q_{\infty}=-1$, whereas AdS spacetime belongs to a distinct class with $Q_{\infty}=0$, is also relevant to the flat-space limit of AdS spacetime. From the perspective of our classification, the AdS sector remains distinct from the Minkowski/dS sector, even as the curvature of AdS can be made arbitrarily small in the limit $\Lambda\rightarrow0^-$ because
\begin{eqnarray}
\lim_{\Lambda\rightarrow0^-}Q_\infty(\text{AdS})=0,    
\end{eqnarray}
whereas
\begin{eqnarray}
Q_\infty(\text{Minkowski})=-1.    
\end{eqnarray}
Within the present classification, this indicates that the flat-space limit of AdS spacetime is nontrivial. This observation is interesting in light of the AdS distance conjecture \cite{Lust2019,Nam2023}, which associates the $\Lambda\rightarrow0^-$ limit of the AdS with an infinite tower of increasingly light states, breaking the low-energy description, and hence this limit belongs to the Swampland \cite{Palti2019}. In this sense, the distinction identified by $Q_{\infty}$ may provide a complementary geometric perspective on the nontrivial nature of the flat-space limit of the AdS spacetime. We stress, however, that the AdS distance conjecture is not derived from the present analysis, and the asymptotic topological flux $Q_\infty$ does not by itself imply an obstruction to the $\Lambda\rightarrow0^-$ limit.

Finally, we present a comparison between our present work and the Wei-Liu-Man construction \cite{SWWei2022}, which provides the principal conceptual differences, mathematical ingredients, and physical outcomes of the two approaches. This comparison is shown in Table \ref{comparis}.
\begin{table}[!htp]
\centering
\begin{tabular}{|c|c|c|}
  \hline
 Feauture & Wei-Liu-Mann construction & Present work \\
  \hline
 Thermodynamic potential  & Off-shell Gibbs free energy & Off-shell grand free energy \\
 & $\mathcal{F}=M-\frac{S}{\tau}$ & $\hat{\mathcal{F}}=M-\frac{S}{\tau_1}-\sum^{N}_{m=2}\frac{\Tilde{Q}_m}{\tau_m}$\\
  \hline
 Auxiliary variable & $\Theta\in[0,\pi]$ & None \\ 
 \hline
Vector field $\Vec{\phi}$ & $(\partial_{r_h}\mathcal{F},-\cot{\Theta}\csc\Theta)$ & Gradient of $\hat{\mathcal{F}}$ with respect to  \\ 
& & thermodynamic variables $(S,\Tilde{Q}_2,...,\Tilde{Q}_{N})$\\
 \hline
Topological map &  $\Vec{n}: (r_h,\Theta,\tau)\setminus Z(\phi)\longrightarrow S^1$ &  $\Vec{n}: \mathcal{M}^{(2N)}\setminus Z(\phi)\longrightarrow S^{N-1}$ \\ 
\hline
Topological current &  One-index topological current $j^\mu$ & Higher-rank topological tensor $j^{\nu_1...\nu_N}$ \\ \hline
Local quantity & Winding number $\omega_i$& Index/degree $\text{index}_{x_i}(\Vec{\phi})$  \\ 
&  of an isolated zero & of an isolated zero \\ \hline

Global quantity & The total winding number: & Topological flux:   \\ 
 & $W=\sum_i\omega_i$ & $Q_{\{\tau_m\}}=\sum_{x_i\in\Sigma^{(N)}_{\{\tau_m\}}}\text{index}_{x_i}(\Vec{\phi})$ \\
\hline

Asymptotic behavior & $W$ depends on the asymptotic & $Q_\infty$ is associated with \\  

&  behavior of black hole temperature & background spacetime geometry \\ \hline

Classification & Black hole thermodynamic defects & $Q_\infty$ is proposed as an invariant \\
& and their stable/unstable branches &  characterizing spacetime backgrounds \\ \hline
\end{tabular}
\caption{A comparison between the Wei-Liu-Mann construction \cite{SWWei2022} and the present work.}\label{comparis}
\end{table}

\section{\label{sec:conclu} Conclusions}

The previous approach to the thermodynamic topology of black holes \cite{SWWei2022} introduces an auxiliary variable $\Theta\in[0,\pi]$ to construct the relevant vector field whose zeros identify the topological/thermodynamic defects corresponding to the black hole solutions. In the present work, we develop an alternative framework for investigating black hole thermodynamics in thermal and chemical equilibrium with the surrounding environment, without introducing such an auxiliary variable. Specifically, the gradient of the off-shell grand free energy with respect to the thermodynamic state variables defines a higher-dimensional vector field. This construction naturally leads to a conserved higher-rank topological tensor, whose flux is determined by the sum of the local topological indices of the relevant zeros. The resulting flux provides a global topological characterization of the thermodynamic defect structure. At each zero, the local topological index characterizes the thermodynamic properties of the corresponding defect and, in particular, distinguishes locally stable and unstable black hole solutions. 

We define the asymptotic topological flux as the asymptotic value of the topological flux carried by the defects contained within the finite hypersurfaces that are continuously pushed toward the asymptotic boundary of the ensemble parameter space. In this limit, the black hole geometry approaches the associated asymptotic/background spacetime geometry. Consequently, the asymptotic topological flux is expected to be determined solely by the asymptotic/background spacetime geometry. We therefore conjecture that the asymptotic topological flux defines an invariant associated with the asymptotic/background spacetime geometry.

Based on the asymptotic topological flux, we have classified maximally symmetric constant-curvature spacetimes into two distinct classes: the first class consists of the Minkowski and dS spacetimes, whereas the AdS spacetime belongs to a different class. This classification suggests that, although the curvature of AdS can approach zero in the limit $\Lambda\rightarrow0^{-}$, the corresponding asymptotic class does not approach that of Minkowski spacetime. Thus, within the present framework, the flat-space limit of AdS is nontrivial. This result is consistent with the fact that the $\Lambda\rightarrow0^{-}$ limit is expected to be a nontrivial limit from the perspective of the AdS distance conjecture \cite{Lust2019,Nam2023}, which associates this limit with the emergence of an infinite tower of increasingly light states and, consequently, a breakdown of the low-energy effective description. Our result may therefore provide a complementary perspective on the nontrivial nature of the flat-space limit of the AdS spacetime.

\section*{Acknowledgments}
We would like to express sincere gratitude to the Referee for his/her constructive comments, suggestions, and questions, which have helped us to clarify important points and hence improve the quality of the paper. We appreciate JINR (Dubna) for the hospitality where part of this work was done. This work is funded by the Scientific Co-operation between DFG and NAFOSTED under grant number DFG.103-2024.01.

\section*{\label{D3-append} Appendix: The mathematical origin of the topological tensor $j^{\nu_1\ldots\nu_N}$}

The topological tensor $j^{\nu_1\ldots\nu_N}$ arises naturally from the topology of the normalized vector field $\Vec{n}=\Vec{\phi}/||\Vec{\phi}||$. This vector field, which satisfies $\Vec{n}^2=1$, takes values on a unit $(N-1)$-sphere $S^{N-1}$. Away from the zero set $Z(\phi)$ of $\Vec{\phi}$, the normalized vector field defines a map as follows
\begin{eqnarray}
\Vec{n}: \mathcal{M}^{(2N)}\setminus Z(\phi)\longrightarrow S^{N-1}.   
\end{eqnarray}
For a zero $x_i\in Z(\phi)$, we choose a sufficiently small $N$-dimensional ball $\mathcal{B}_i$ transverse to $Z(\phi)$ such that $\mathcal{B}_i\cap Z(\phi)=\{x_i\}$ with the boundary being an $(N-1)$-sphere $\mathcal{S}_i^{N-1}=\partial\mathcal{B}_i$. On this boundary, the normalized vector field $\Vec{n}$ is regular and hence it defines the following map
\begin{equation}
\Vec{n}:\mathcal{S}_i^{N-1}\longrightarrow S^{N-1}.
\end{equation}
The topological index associated with the zero $x_i$ is the degree of this map, given by
\begin{equation}
\text{index}_{x_i}(\Vec{\phi})=\int_{\mathcal{S}_i^{N-1}} n^*\Omega_{S^{N-1}},
\end{equation}
where $\Omega_{S^{N-1}}$ is the normalized volume form on $S^{N-1}$, satisfying $\int_{S^{N-1}}\Omega_{S^{N-1}}=1$, written as
\begin{eqnarray}
\Omega_{S^{N-1}}=\frac{1}{(N-1)!\,\omega_{N-1}}
\epsilon_{a_1\ldots a_N} n^{a_1} dn^{a_2}\wedge\ldots\wedge dn^{a_N},
\end{eqnarray}
with $\omega_{N-1}$ denoting the volume of the unit $S^{N-1}$, and $n^*\Omega_{S^{N-1}}$ denotes the pullback of $\Omega_{S^{N-1}}$. By Stokes' theorem, we find
\begin{eqnarray}
\text{index}_{x_i}(\Vec{\phi})=\int_{\mathcal{B}_i}d\left(n^*\Omega_{S^{N-1}}\right).
\end{eqnarray}
Since $\Omega_{S^{N-1}}$ is closed, $d(n^*\Omega_{S^{N-1}})=0$ on the defect-free region. Its nonvanishing contribution is therefore localized on the zero set $Z(\phi)$, where $\Vec{n}$ becomes singular. Accordingly, $d(n^*\Omega_{S^{N-1}})$ is understood as a distributional $N$-form. The corresponding conserved topological tensor is obtained by taking the Hodge dual of this distributional $N$-form as follows
\begin{equation}
j=\star d\!\left(n^*\Omega_{S^{N-1}}\right),
\end{equation}
whose components are the antisymmetric tensor $j^{\nu_1\ldots\nu_N}$. For $\mathcal{B}_i$ taken at fixed ensemble variables $(\tau_1,\ldots,\tau_N)$, the flux of the topological tensor through $\mathcal{B}_i$ is therefore
\begin{eqnarray}
\int_{\mathcal{B}_i}j^{\tau_1\ldots\tau_N}d\Sigma_{\tau_1\ldots\tau_N}=\int_{\mathcal{B}_i}d\left(n^*\Omega_{S^{N-1}}\right)=\text{index}_{x_i}(\Vec{\phi}).   
\end{eqnarray}
In this way, the flux of $j^{\nu_1\ldots\nu_N}$ through a transverse surface measures the index of the normalized map around the corresponding defect. This provides the geometric and topological origin of the tensors introduced in Eqs. (\ref{simp-topo-tens}) and (\ref{gen-topo-tens}).

\end{document}